\documentclass[twocolumn,aps,prb]{revtex4}
\usepackage[T1]{fontenc}
\usepackage[latin1]{inputenc}
\usepackage{amsmath}
\usepackage{amstext}
\usepackage{epsfig}
\usepackage{epic}
\usepackage{eepic}
\usepackage{amssymb}
\usepackage{exscale}
\usepackage{pst-node,array}
\begin{document}
\title{A renormalized excitonic method in terms of block excitations. Application to spin lattices}
\author{Mohamad Al Hajj, Jean-Paul Malrieu, and Nathalie Guih\'ery}
\affiliation{Laboratoire de Physique Quantique, IRSAMC/UMR5626, Universit\'e Paul Sabatier, 118 route 
de Narbonne, F-31062 Toulouse Cedex 4, France}
\begin{abstract}
Dividing the lattice into blocks with singlet ground state and knowing the exact low energy spectrum
of the blocks and of dimers (or trimers) of blocks, it is possible to approach the lowest part of the 
lattice spectrum through an excitonic type effective model. The potentialities of the method are
illustrated on the 1-D frustrated chain and the 1/5-depleted square and the plaquette 2-D lattices. The method
correctly locates the phase transitions between gapped and non-gapped phases.
\bigskip
\end{abstract}
\maketitle
\section{Introduction}
The idea that one may see a periodic lattice as built from interacting blocks of sites, rather than as 
interacting sites, is computationally and intellectually attractive and has received several exploitations. This process 
consists in a scale change. Wilson's bright proposal of real space renormalization group (RSRG)\cite{Ref1}
is certainly the most elegant illustration of this idea, since it can be infinitely iterated and 
asymptotically converges on physically meaningful accumulation points. 
The method in its original version consists in a simple and severe truncation of the Hilbert 
space by considering only the lowest states of each block
and the products of the selected block eigenfunctions to approach the  eigenfunctions of blocks of blocks. 
The method  happened to be numerically inefficient. 
Recent works have shown that its performances can be dramatically improved when one defines effective
interactions between the blocks. These effective interactions are calculated from the knowledge
of the exact spectrum of the dimers (or trimers) of blocks using Bloch's theory\cite{Ref2} of effective 
Hamiltonian. 
The so-called contractor renormalization (CORE) technique \cite{Ref3,Ref4,Ref5,Ref6}
considers blocks, retains a few eigenstates $\vert I_A\rangle$ of
each block $A$ and defines inter-block effective interactions
$\langle I_A J_B \vert H^{eff} \vert K_A L_B \rangle$, using the spectrum of the $AB$ problem and the
Bloch's theory\cite{Ref2} of effective Hamiltonians.
In most applications the size of the blocks remains small, several states per blocks are kept, and three and / or 
four blocks effective interactions are introduced from the knowledge of the spectrum of trimers and / or tetramers
of blocks. A specific variant of the method has been proposed by two of the authors under the name of RSRG-EI 
\cite{Ref7,Ref8,Ref9} (RSRG with effective interactions) which treats spin lattices by considering blocks with an odd 
number of sites and a doublet ground state which is the only one to be explicitily retained. The blocks are then quasi-spin. From the spectrum of dimers or trimers of 
blocks one may define an inter-block Heisenberg Hamiltonian. A proper design of the blocks frequently results in an 
isomorphism between the original lattice and the lattice of blocks. Hence the process may be iterated, exhibiting critical 
ratios of the elementary interactions and accumulation points. The methodological studies have examined the following 
dilemma in the research of accuracy
\begin{itemize} 
\item[-] consider larger blocks and only dimers of blocks (i.e., two-body effective interactions only) or 
\item[-] consider smaller blocks and trimers or tetramers (i.e.,three and four blocks interactions)
\end{itemize}
showing that in most cases the former solution is sufficient.
In a rather similar spirit one may mention the block 
correlated coupled cluster method\cite{Ref10} which employs the coupled cluster formalism 
\cite{Ref11,Ref12,Ref13} starting from the product of the ground 
state eigenfunctions for each block. A similar philosophy (with small blocks) is present in the 
applications of self-consistent perturbation formalism to periodic lattices.\cite{Ref14} These last 
two methods
do not provide information on the gaps, while CORE and RSRG-EI bring good estimates of them. 

The present work is closely related but is focused on a direct research of the gap. One starts now
from blocks constituted of an even number of sites and having a non-degenerate singlet ground state. 
The blocks may be identical or not, but they must lead to a periodic picture of the lattice
in terms of blocks (hence with larger unit cells). The ground state will be built from products of 
block ground states. Considering the exact energies of dimers or trimers of blocks, one will define
effective interactions between blocks in their ground states, producing an additive energy systematics.

For the study of excited states, one will also consider the lowest states of the blocks and of the 
various dimers and eventually trimers of blocks. The knowledge of the lowest states of the dimers and trimers
(energy and eigenvectors) enables 
one to define the effective interactions between an excited block and neighbor ground state blocks
and effective excitation transfer integrals from one block to its neighbors. These quantities will
be defined through the Bloch's theory of effective Hamiltonians. Then the lattice lowest excitations
are treated through an excitonic model that makes use of these effective quantities. The theory is 
developed in section II. Section III first shows the improvement brought by the use of effective 
interactions rather than of bare interactions on elementary mono-electronic problems where
the original RSRG version failed. The efficiency of the method will then be illustrated on three spin
lattices, namely the 1-D frustrated chain, the 1/5-depleted 2-D square lattice and the plaquette lattice. 
The three problems exhibit phase transitions (of second and first order) which are 
satisfactorily treated with the here-proposed renormalized excitonic method (REM).
\section{Method}
\subsection{Principle}
Let us consider a lattice constituted of blocks $A,B...$ having a non-degenerate singlet ground state.
$\psi^{0}_{A}$ is the ground state for the block $A$, of Hamiltonian $H_A$, 
\begin{equation}
H_A \psi^{0}_{A} = E^{0}_{A} \psi^{0}_{A}. 
\end{equation}
The zero-order description of the \textit{ground state} of the lattice will be the 
product of the ground states for each block 
\begin{equation}
\Psi_0=\prod_A \psi^{0}_{A}.
\label{eq2}
\end{equation}
The zero-order energy would be additive
\begin{equation}
E^0=\sum_A E^{0}_{A},
\end{equation}
but since the Hamiltonian involves interaction operators between blocks
\begin{equation}
H=\sum_A H_{A}+\sum_{A} \sum_{<B} V_{AB},
\end{equation}
the mean energy implies inter-block interactions
\begin{eqnarray}
\langle \Psi_0 \vert H \vert \Psi_0 \rangle & = & \sum_A E^{0}_{A}+\sum_{A}\sum_{<B} \langle \psi^{0}_{A} 
\psi^{0}_{B} \vert V_{AB} \vert \psi^{0}_{A}\psi^{0}_{B} \rangle \nonumber \\
& = & \sum_A E^{0}_{A}+\sum_{A} \sum_{<B} v^{0}_{AB}.
\end{eqnarray}
Here $v^{0}_{AB}$ is a zero-order interaction energy between blocks $A$ and $B$. 

Solving exactly the $AB$ problem
\begin{equation}
(H^{0}_{A}+H^{0}_{B}+V_{AB})\vert \Psi^{0}_{AB}\rangle = E^{0}_{AB} \vert \Psi^{0}_{AB}\rangle,
\end{equation}
enables one to define an improved interaction energy $v_{AB}$ 
\begin{equation}
v_{AB}=E^{0}_{AB}-E^{0}_{A}-E^{0}_{B},
\end{equation}
which takes into account at all orders the perturbative effect of excitations on $A$ and $B$, i.e., of the
vectors $\vert \psi_A^{i}\psi_B^{j}\rangle.$ Going to trimers it would be possible to define a quantity
\begin{equation}
v_{ABC}=E^{0}_{ABC}-E^{0}_{A}-E^{0}_{B}-E^{0}_{C}-v_{AB}-v_{BC}-v_{AC}.
\end{equation}
The ground state energy will be additive 
\begin{equation}
E=\sum_A E^{0}_{A}+\sum_{A} \sum_{<B} v_{AB}+\sum_{A}\sum_{<B} \sum_{<C} v_{ABC}+ \cdots
\end{equation}
At this stage we would like to stress on the fact that the treatment of the ground state is from a single function 
$\Psi_0$ given by (Eq. \ref{eq2}). 

For the description of the \textit{excited states} we shall consider a very limited model space.
If one considers the lowest excited state on $A$, $\psi_A^{\ast}$, of the desired spin multiplicity,
\begin{eqnarray}
H_A\vert\psi_A^{\ast}\rangle & = & E^{\ast}_{A}\vert\psi_A^{\ast}\rangle, \\
\vert\psi_A^{\ast}\rangle & = & T_A^+\vert\psi_A^{0}\rangle, \nonumber
\end{eqnarray}
the on-block excitation lowest energy is $\Delta^{\ast}_{A}=E^{\ast}_{A}-E^{0}_{A}.$
We intend to represent the lowest excitations on an ensemble of blocks from localy singly excited configurations of 
the type $\Psi_I^{\ast}=T_I^+\Psi^0=\vert\psi_A^0 \cdots \psi_H^0\psi_I^{\ast}\psi_J^0 \cdots \vert$ 
where the excitation is localized on block $I$. This will lead to an excitonic treatment of the excitation
\begin{equation}
\Psi^{\ast}=\sum_{I}\lambda_I \Psi_I^{\ast}=\left(\sum_{I}\lambda_IT_I^+\right)\Psi^{0}.
\label{eq11}
\end{equation}
In order to establish the corresponding model Hamiltonian one needs to calculate
\begin{itemize}
\item[-] the effective interaction between an excited block and neighbor blocks in their ground state
\item[-] the excitation hopping integrals which propagate an excitation from one block to other blocks.
\end{itemize}
These informations will be obtained from the spectral properties of dimers and trimers of blocks, using the 
effective Hamiltonian theory. Let us consider first the extraction of information from interacting pairs of blocks.
\subsection{Extraction of information from dimers of blocks}
For a dimer $AB$ one may chose as model space for the lowest energy excited states the two vectors 
$\psi_A^{\ast} \psi_B^{0}$ and $\psi_A^{0} \psi_B^{\ast}.$ The corresponding projector is 
\begin{eqnarray}
P_{AB}^{\ast} & = & \vert\psi_A^{\ast}\psi_B^{0}\rangle\langle\psi_A^{\ast}\psi_B^{0}\vert +
\vert\psi_A^{0}\psi_B^{\ast}\rangle\langle\psi_A^{0}\psi_B^{\ast}\vert \\ \nonumber
&= & \vert T_A^+\Psi^{0}\rangle\langle T_A^+\Psi^{0}\vert+\vert T_B^+\Psi^{0}\rangle\langle T_B^+\Psi^{0}\vert.
\end{eqnarray}
If one identifies the two eigenvectors $\Psi_{AB}^{\ast}$ and $\Psi_{AB}^{\ast'}$ of eigenenergies
$E^{\ast}_{AB}$ and $E^{\ast'}_{AB}$ 
\begin{eqnarray}
H_{AB}\vert\Psi_{AB}^{\ast}\rangle & = & E^{\ast}_{AB}\vert\Psi_{AB}^{\ast}\rangle, \\ 
H_{AB}\vert\Psi_{AB}^{\ast'}\rangle & = & E^{\ast'}_{AB}\vert\Psi_{AB}^{\ast'}\rangle,
\end{eqnarray}
which have the largest projections onto the model space, it 
is possible to define an effective Hamiltonian built on the model space and according to Bloch's 
definition 
\begin{equation}
H^{eff}\vert P_{AB}^{\ast} \Psi_{AB}^{\ast} \rangle =E^{\ast}_{AB} \vert P_{AB}^{\ast} 
\Psi_{AB}^{\ast} \rangle,
\end{equation}
\begin{equation}
H^{eff}\vert P_{AB}^{\ast} \Psi_{AB}^{\ast'} \rangle =E^{\ast'}_{AB} \vert P_{AB}^{\ast}
\Psi_{AB}^{\ast'} \rangle.
\end{equation}
In order to have an hermitian effective Hamiltonian its eigenvectors must be orthogonal. We shall
assume that $P_{AB}^{\ast} \vert \Psi_{AB}^{\ast'}\rangle$ is orthogonal or Schmidt-orthogonalized to 
$P_{AB}^{\ast} \vert \Psi_{AB}^{\ast}\rangle.$ One may write, after normalization,
\begin{equation}
\vert P_{AB}^{\ast} \Psi_{AB}^{\ast} \rangle =a \vert \psi_A^{\ast} \psi_B^{0}\rangle + b 
\vert \psi_A^{0} \psi_B^{\ast}\rangle,
\end{equation}
\begin{equation}
\vert P_{AB}^{\ast} \Psi_{AB}^{\ast'} \rangle =-b \vert \psi_A^{\ast} \psi_B^{0}\rangle + a
\vert \psi_A^{0} \psi_B^{\ast}\rangle.
\end{equation}
The spectral definition of $H^{eff}$ leads to the following equations
\begin{eqnarray}
\langle \psi_{A}^{\ast}\psi_B^{0} \vert H^{eff}\vert \psi_{A}^{\ast}\psi_B^{0} \rangle & = &
a^2 E^{\ast}_{AB}+b^2 E^{\ast'}_{AB} \nonumber \\
& = &  E^{\ast}_{A}+E^{0}_{B}+v_{(A^{\ast})B},
\end{eqnarray}
\begin{eqnarray}
\langle \psi_{A}^{0}\psi_B^{\ast} \vert H^{eff}\vert \psi_{A}^{0}\psi_B^{\ast} \rangle & = &
b^2 E^{\ast}_{AB}+a^2 E^{\ast'}_{AB} \nonumber \\
& = & E^{0}_{A}+E^{\ast}_{B}+v_{A(B^{\ast})},
\end{eqnarray}
\begin{equation}
\langle \psi_{A}^{\ast}\psi_B^{0} \vert H^{eff}\vert \psi_{A}^{0}\psi_B^{\ast} \rangle =
(E^{\ast}_{AB}-E^{\ast'}_{AB})ab = h_{AB}.
\end{equation}
The terms $v_{(A^{\ast})B}$ (resp. $v_{A(B^{\ast})}$) represent the effective interactions between 
$A^{\ast}$ and $B$ (resp. between $A$ and $B^{\ast}$) and $h_{AB}$ is the effective interaction
responsible for the transfer of excitation from $A$ to $B$. If $A$ and $B$ are identical blocks and 
if the $AB$ dimer presents an element of symmetry transforming $A$ in to $B$ and vice versa, 
$\vert a \vert =\vert b\vert=1/\sqrt{2}$, one eigenvector is an in-phase combination of 
$\psi_{A}^{\ast}\psi_B^{0}$ and $\psi_{A}^{0}\psi_B^{\ast}$, of energy $E^{\ast g}_{AB}$, the other one being the 
out-of-phase combination, of energy $E^{\ast u}_{AB}$.
\begin{eqnarray}
P_{AB}^{\ast} \Psi_{AB}^{\ast g} & = & \frac{1}{\sqrt{2}}(\psi_{A}^{\ast}\psi_B^{0}+\psi_{A}^{0}\psi_B^{\ast}), \\ 
H_{AB} \vert\Psi_{AB}^{\ast g}\rangle & = & E^{\ast g}_{AB} \vert\Psi_{AB}^{\ast g}\rangle, 
\end{eqnarray}
\begin{eqnarray}
P_{AB}^{\ast} \Psi_{AB}^{\ast u} & = & \frac{1}{\sqrt{2}}(\psi_{A}^{\ast}\psi_B^{0}-\psi_{A}^{0}\psi_B^{\ast}), \\ 
H_{AB} \vert\Psi_{AB}^{\ast u}\rangle & = & E^{\ast u}_{AB} \vert\Psi_{AB}^{\ast u}\rangle, 
\end{eqnarray}
\begin{equation}
v_{(A^{\ast})B}=v_{A(B^{\ast})}=\frac{1}{2}(E^{\ast g}_{AB}+E^{\ast u}_{AB})-E^{\ast}_{A}-E^{0}_{B},
\end{equation}
\begin{equation}
h_{AB}=\frac{1}{2}(E^{\ast g}_{AB}-E^{\ast u}_{AB}).
\end{equation}
It is then possible to consider the infinite lattice in which each block is surrounded by
nearest-neighbor blocks $B$ with equal or different respective interactions. The ground state being 
represented by $\vert \underset{K}{\prod} \psi_K^{0} \vert$ has an energy 
\begin{equation}
\mathcal{E}_{0}=\langle\Psi^0\vert H\vert\Psi^0\rangle=\sum_K E_K^{0}+ \sum_{K}\sum_{<L}v_{KL}.
\end{equation}
The set of excited states described by our method will be built in the space spanned by all products
\begin{equation}
\Psi_I^{\ast}=T_I^+\Psi^0=
\left(
\prod_{K=0,I-1}\psi_K^{0}
\right)
\psi_I^{\ast}
\left(
\prod_{L=I+1,\infty}\psi_L^{0}
\right).
\end{equation}
Their effective energy is 
\begin{eqnarray}
\mathcal{E}_{I}^{\ast} & = & \langle \Psi_I^{\ast} \vert H^{eff}\vert \Psi_I^{\ast} \rangle 
 =  \sum_{K\not=I} E_K^{0}+E_{I}^{\ast}
\nonumber  \\
& + & \sum_{K(\not=I)}\sum_{<L(\not=I)} v_{KL}+ \sum_{K\not=I}  v_{K(I^{\ast})}.
\end{eqnarray}
The local excitation energy is 
\begin{equation}
\mathcal{E}_{I}^{\ast} - \mathcal{E}_{0} = E_{I}^{\ast}-E_{I}^{0} + \sum_{K\not=I}(v_{K(I^{\ast})}- v_{KI}).
\end{equation}
The vectors $\Psi_I^{\ast}$ and $\Psi_J^{\ast}$ interact through the matrix element 
$\langle T_I^+\Psi^0\vert H^{eff}\vert T_J^+\Psi^0\rangle=\langle\Psi_I^{\ast}\vert H^{eff}\vert\Psi_J^{\ast}\rangle =h_{IJ}.$
The effective Hamiltonian matrix has a near-diagonal structure, similar to that of a tight-binding 
mono-electronic Hamiltonian. It generates bands which only 
represent the states of the lattice having large projections onto the vectors $\Psi_I^{\ast}$, i.e.,
on the intra-blocks lowest energy excitations. The descriptions of the lowest energy states of the
lattice should be relevant.
If the blocks are identical and engaged in the same interactions of negative sign with their first 
neighbors, the excitation energy to the lowest $\overrightarrow{k}=0$  state should be 
\begin{equation}
\Delta^{\ast\infty}=(E_{I}^{\ast}-E_{I}^{0})+\sum_{K\not=I}(v_{K(I^{\ast})}- v_{KI})+
\sum_{K\not=I} h_{IK}.
\label{eq26}
\end{equation}
If the $KI$ couples present an element of symmetry transforming $K$ into $I$, using Eq. \ref{eq26} and one obtains,
\begin{equation}
\Delta^{\ast\infty}=\Delta^{\ast}_{I}+\sum_{K}(\Delta^{\ast}_{KI}-\Delta^{\ast}_{I})
\label{eq27}
\end{equation}
where $\Delta^{\ast}_{I}=E_{I}^{\ast}-E_{I}^{0}$, $\Delta^{\ast}_{KI}=E_{KI}^{\ast g}-E_{KI}^{0}$
are excitation energies on the blocks and dimers of blocks respectively. One notices that the 
other root $E_{KI}^{\ast u}$ of the dimer disappears in this expression. 

One sees that the derivation leads to a renormalized excitonic method, where the excitation transfer
integrals $h_{IJ}$ are renormalized, therefore including to all orders some indirect processes
going through higher-energy (multiple) excitations on neighbor blocks or inter-block excitations as 
will be shown hereafter. Of course the results will be dependent on the shape and size $n$ of the blocks.
\subsection{Extraction of information from trimers of blocks}
It is possible to use the eigenstates of trimers of blocks to extract three blocks interactions. For a given
shape of the elementary blocks one must of course consider the various types of trimers of blocks. While for the
ground state the three-block correction is given by
\begin{equation}
v_{ABC}=E^0_{ABC}-E^0_{A}-E^0_{B}-E^0_{C}-v_{AB}-v_{AC}-v_{BC},
\end{equation}
for the excited states the model space involves three vectors. The projector on the model space is  
\begin{eqnarray}
P^{\ast}_{ABC} & = & \vert \psi_A^{\ast}\psi_B^{0}\psi_C^{0}\rangle \langle \psi_A^{\ast}\psi_B^{0}\psi_C^{0} \vert
+ \vert \psi_A^{0}\psi_B^{\ast}\psi_C^{0}\rangle \langle \psi_A^{0}\psi_B^{\ast}\psi_C^{0} \vert
\nonumber \\ 
& & +\vert \psi_A^{0}\psi_B^{0}\psi_C^{\ast} \rangle \langle \psi_A^{0}\psi_B^{0}\psi_C^{\ast} \vert \nonumber \\
& = &\vert T_A^+\Psi^0\rangle\langle T_A^+\Psi^0\vert+\vert T_B^+\Psi^0\rangle\langle T_B^+\Psi^0\vert \nonumber \\
& & +\vert T_C^+\Psi^0\rangle\langle T_C^+\Psi^0\vert,
\end{eqnarray}
one must identify the three eigenstates of the $ABC$ problem having the largest projections on the model space. 
Notice that these three states are not necessarily the three lowest ones. Let us call 
$\Psi^{\ast}_{ABC}$, $\Psi^{\ast'}_{ABC}$ and $\Psi^{\ast''}_{ABC}$ these three states and 
$E^{\ast}_{ABC}$, $E^{\ast'}_{ABC}$ and $E^{\ast''}_{ABC}$ the corresponding eigenenergies. For hermiticity the 
projected eigenvectors are orthonormalized leading to three vectors $\Phi^{\ast}_{ABC}=P^{\ast}_{ABC}\Psi^{\ast}_{ABC}$, 
$\Phi^{\ast'}_{ABC}=P^{\ast}_{ABC}\Psi^{\ast'}_{ABC}$ and 
$\Phi^{\ast''}_{ABC}=\Psi^{\ast''}_{ABC}P^{\ast}_{ABC}$ and from the spectral definition of $H^{eff}$  
\begin{eqnarray}
H^{eff}_{ABC} & = & E^{\ast}_{ABC} \vert \Phi^{\ast}_{ABC}\rangle \langle \Phi^{\ast}_{ABC} \vert+
E^{\ast'}_{ABC} \vert \Phi^{\ast'}_{ABC} \rangle \langle \Phi^{\ast'}_{ABC} \vert 
\nonumber \\
& & +E^{\ast''}_{ABC} \vert \Phi^{\ast''}_{ABC} \rangle \langle \Phi^{\ast''}_{ABC} \vert,
\end{eqnarray}
one may calculate the diagonal matrix elements of $H^{eff}_{ABC}$ and reexpress its matrix elements as 
\begin{eqnarray}
\langle \psi_A^{\ast}\psi_B^{0}\psi_C^{0} \vert H^{eff}_{ABC} \vert \psi_A^{\ast}\psi_B^{0}\psi_C^{0} \rangle  =  
E^{\ast}_{A}+E^0_{B}+E^0_{C}+
\nonumber \\
v_{(A^{\ast})B}+v_{(A^{\ast})C}+v_{BC}+v_{(A^{\ast})BC},
\end{eqnarray}
which defines a three-body interaction $v_{(A^{\ast})BC}$, and revised excitation hopping integrals
\begin{equation}
\langle \psi_A^{\ast}\psi_B^{0}\psi_C^{0} \vert H^{eff}_{ABC} \vert \psi_A^{0}\psi_B^{\ast}\psi_C^{0} \rangle =
h_{AB}+h_{AB(C)}.
\end{equation}
The last term represents the effect of $C$ on the hopping between $A$ and $B$. One also obtains effective hopping between
non directly interacting blocks (for instance $A$ and $C$ through $B$ in a linear $ABC$ configuration). This indirect
propagation may proceed, for instance for triplet states, through the process 
$\psi_A^{\ast}\psi_B^{0}\psi_C^{0} \longleftrightarrow \psi_A^{\ast}\psi_B^{\ast}\psi_C^{\ast} \longleftrightarrow
\psi_A^{0}\psi_B^{0}\psi_C^{\ast}.$
These effective interactions are used in the excitonic treatment. 

For the periodic lattice
\begin{widetext}
\begin{eqnarray}
\langle T^+_I\Psi^0 \vert H^{eff}\vert T^+_I\Psi^0\rangle & = &  
\langle\cdots\psi_H^{0}\psi_I^{\ast}\psi_J^{0}\cdots\vert H^{eff}\vert\cdots\psi_H^{0}\psi_I^{\ast}\psi_J^{0} 
\cdots\rangle = \sum_{J\not=I} E_J^{0}+E^{\ast}_{I} 
+\sum_{J(\not=I)}\sum_{<K(\not=I)}v_{JK}
\nonumber \\
& & +\sum_{J\not=I}v_{(I^{\ast})J} 
+\sum_{J(\not=I)}\sum_{<K(\not=I)}\sum_{<L(\not=I)}v_{JKL}
+\sum_{J(\not=I)}\sum_{<K(\not=I)}v_{(I^{\ast})JK}, 
\\
\langle T^+_I\Psi^0 \vert H^{eff}\vert T^+_J\Psi^0\rangle & = & 
\langle \cdots \psi_H^{0}\psi_I^{\ast}\psi_J^{0} \cdots \vert H^{eff} \vert \cdots \psi_H^{0}\psi_I^{0}\psi_J^{\ast} 
\cdots \rangle  = 
h_{IJ}+\sum_{K} h_{IJ(K)}.
\end{eqnarray}
\end{widetext}
The excitation energy for the vector $\overrightarrow{k}=0$ (which is not necessarily the lowest one), when all
blocks are equivalent, is 
\begin{eqnarray}
\Delta^{\ast\infty}  = E^{\ast}_{I}-E^{0}_{I}+\sum_{J}(v_{(I^{\ast})J}-v_{IJ})+
\nonumber \\
\sum_{JK}(v_{(I^{\ast})JK}-v_{IJK})+\sum_{J}(h_{IJ}+\sum_{K} h_{IJ(K)}).
\end{eqnarray}
The method is generalizable to four (and more) blocks. One should however remark that when one increases the number of 
blocks the identification of the eigenstates having the largest projections onto the model space may become ambiguous. 
When changing the ratio of the intersite interactions the (say) 3rd best vector may jump from the 
eigenvector number 3 to the eigenvector number 4, a problem which will be documented below. In such a case the 
effective Hamiltonian will be a discontinuous function of the intersite interactions, which is a rather 
unpleasant feature. 
\subsection{Comment}
Of course the method is anly applicable to the study of gapped systems and to locate the phase transition between a 
gapped phase and a gapless phase. This limit is due to the fact that one uses different model spaces for the ground 
state and for the excited states. The method cannot provide the low energy physics of gapless anti-ferrmagnetic lattices.
For such phases the method is unable to give a strictly zero gap nor the density of states. As will be shown in the 
following examples the calculated gap becomes extremely small and in some cases it may be spuriously negative. This
limit (which is not present in CORE method) should be kept in mind.
\section{Test applications}
\subsection{Illustration of the superiority of effective interactions over bare ones}
This first subsection will illustrate the crucial effect of considering effective interactions in 
excitonic models. We shall consider textbook problems concerning the 1-D non-dimerized chain using a 
tight-binding mono-electronic Hamiltonian. 
\begin{equation}
H=t\sum_{\langle p, q \rangle}(a_{p}^{+}a_{q}+a_{q}^{+}a_{p}).
\end{equation}

We shall address successively three problems, namely, 
\begin{itemize}
\item[-] the position of the highest orbital for half filling, 
\item[-] the energy of the lowest occupied orbital, and
\item[-] the size-dependence of the lowest excitation energy at half filling.
\end{itemize}
In the three problems we shall demonstrate that a bare excitonic model gives an incorrect behavior of the 
properties, introducing spurious dependence on the size of the blocks. On the contrary the definition of 
effective interactions between the blocks dramatically reduces the error coming from the use of finite size 
blocks.

One may address first the question of the Fermi level for the half-filled band, which is of zero energy.
One shall consider blocks of size $2n$. In each of them (say $I$) the ground state is 
\begin{equation}
\psi_I^0=\prod_{k=1,n}\phi_I^k
\end{equation}
where $H_I\phi_I^k=\mathcal{E}^k\phi_I^k$ with $\mathcal{E}^k=2t\cos\frac{k\pi}{2n+1}$.\\
For the whole system the ground state is treated from 
\begin{equation}
\Psi^0=\prod_{I}\psi_I^0.
\end{equation}
The lowest ionization process in each block concerns the highest occupied orbital $\phi_I^n$,
\begin{equation}
\psi_I^{\ast}=a_{\phi_I^n}\psi_I^0.
\end{equation}
The lowest ionization energy for the block is 
\begin{equation}
\Delta_I^{\ast}=-\mathcal{E}^n=-2t\cos\frac{n\pi}{2n+1}\simeq-\frac{t\pi}{2n+1}+O(2),
\end{equation}
where $O(2)$ is proportional to $n^{-2}$. The ionized state of the whole system will be described according to 
Eq. \ref{eq11} from the vectors $\Psi_I^{\ast}=a_{\phi_I^n}\Psi^0$. The direct matrix element of $H$ between 
$\Psi_I^{\ast}$ and $\Psi_J^{\ast}$ is 
\begin{equation}
\langle a_{\phi_I^n}\Psi^0\vert H\vert a_{\phi_J^n}\Psi^0\rangle=-\langle\phi_I^n \vert H\vert\phi_J^n\rangle=h_{IJ}.
\end{equation}
Since 
\begin{equation}
\vert\phi_{I}^{n}\rangle=\sum_{r}C_n^{r}\vert\chi_{I}^{r}\rangle
\end{equation}
where $\chi_{I}^{r}$ is the orbital localized on site $r$ in block $I$, 
\begin{eqnarray}
h_{IJ}& = & -\langle\phi_{I}^{n}\vert H\vert\phi_{J}^{n}\rangle=-C_n^{2n}C_n^{1}\langle\chi_{I}^{2n}\vert H\vert
\chi_J^{1}\rangle \nonumber \\
& = & \pm(C_n^{1})^2t,
\end{eqnarray}
and since
\begin{equation}
C_n^{r}=\frac{1}{\sqrt{2n+1}}\sin \frac{n\pi r}{2n+1}
\end{equation}
\begin{equation}
h_{IJ}=\frac{t}{2n+1}(\sin \frac{n\pi}{2n+1})^2
=\frac{t}{2n+1}(1+O(2)).
\end{equation}
This is the bare interaction, propagating the hole between adjacent blocks. The bare excitonic 
method predicts therefore the ionization potential of the infinite chain calculated blocks of $2n$ sites as 
\begin{eqnarray}
\Delta^{\ast\infty}(2n) & = &\Delta_I^{\ast}+2h_{IJ} \nonumber \\
& = &-\frac{t\pi}{2n+1}+\frac{2t}{2n+1}=\frac{t(-\pi+2)}{2n+1}.
\end{eqnarray}
This energy is smaller than that of the $2n$ sites block but it remains different from zero and of order $n^{-1}$.
On the contrary, the effective interaction can be calculated from the energies of the highest occupied orbitals 
of the $4n$ sites system. It will lead to a specification of Eq. \ref{eq26}
\begin{eqnarray}
\Delta^{\infty}(n) & = & 2\Delta^{\ast}(4n)-\Delta^{\ast}(2n) \nonumber \\
& = & 2\frac{t\pi}{4n+1}-\frac{t\pi}{2n+1}+O(2)=O(2).
\end{eqnarray}
The spurious $n^{-1}$ dependent term, present in the bare excitonic model, disappears from the effective one, 
as desired. 

As another application one might have looked at the lowest energy level, which should be $2t$ for 
the infinite lattice. For a $2n$ sites block the corresponding energy is 
\begin{equation}
\mathcal{E}^{1}(n)=2t \cos \frac{\pi}{2n+1}=2t(1-\frac{\pi^2}{8n^2}).
\end{equation}
The deviation to the asymptotic value is $\pi^2t/4n^2$. The bare interaction 
\begin{equation}
\langle\phi_{I}^{1} \vert H \vert\phi_{J}^{1} \rangle= \frac{t}{2n+1}\sin^2 \frac{\pi}{2n+1}\simeq 
\frac{t}{n^3}
\end{equation}
\begin{figure}[t]
\centerline{\includegraphics[scale=1.0]{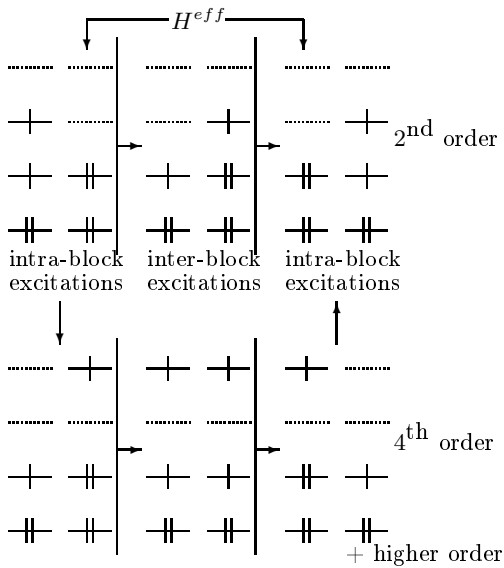}}
\caption{Schematic view of the incorporation of complex effects in the off-diagonal effective
interactions.}
\label{fig1}
\end{figure}
behaves as $n^{-3}$ and cannot compensate the $n^{-2}$ dependent error which affects the 
$\mathcal{E}^{1}(n)$ value. On the contrary the effective interaction between these two blocks is 
$(1/2)(\mathcal{E}^{1}(4n)-\mathcal{E}^{2}(4n))\simeq 3\pi^2t/32n^2$ i.e., has the
expected $n^{-2}$ dependency and reduces the error to $\pi^2t/32n^2$. This problem has an historical
importance since the recognition of the failure of the original RSRG version on this problem 
has led to the development of the density matrix renormalization group (DMRG).\cite{Ref15,Ref16} One sees that the 
consideration of properly extracted effective interactions represents a qualitative improvement
of the RSRG method, as already mentioned elsewhere.\cite{Ref3,Ref4,Ref5,Ref6,Ref7,Ref8} 

As an even more dramatic problem, one may 
consider the excitation gap in the same 1-D problem, at half-filling. In a 2n sites block the local
excitations are ($\phi^n \to \phi^{n+1}$) electron jump around the Fermi level, the corresponding
excitation energy is $\Delta_I^{\ast}=-2t/(2n+1)$. There is no bare interaction between the state which is excited 
on block $I$ and on the one excited on block $J$, due to the mono-electronic nature of $H$, 
\begin{equation}
\langle a_{\phi^{(n+1)}_{I}}^{+}a_{\phi^{n}_{I}}\Psi_0 \vert H \vert a_{\phi^{(n+1)}_{J}}^{+}a_{\phi^{n}_{J}}\Psi_0
\rangle=0.
\end{equation}
Hence the excitation energy would be that of an isolated block and would behave as $n^{-1}$.
On the contrary there is a $n^{-1}$ dependent effective interaction which cancels the spurious $n^{-1}$ term since 
Eq. \ref{eq26} leads to 
\begin{equation}
\Delta^{\ast\infty}=-\frac{2t\pi}{2n+1}+\frac{4t\pi}{4n+1}+O(2)=O(2).
\end{equation}
The term 
\begin{equation}
\langle a_{\phi^{(n+1)}_{I}}^{+}a_{\phi^{n}_{I}}\Psi_0\vert H^{eff}\vert a_{\phi^{(n+1)}_{J}}^{+}a_{\phi^{n}_{J}}
\Psi_0\rangle=\frac{2t\pi}{(4n+1)},
\end{equation}
reflects the indirect interaction through inter-blocks charge transfer states, in particular of the 
type $a_{\phi^{(n+1)}_{I}}^{+}a_{\phi^{n}_{J}}\Psi_0$, which make possible the independent delocalization 
of the hole and of the particle (cf. Fig. \ref{fig1}).
\subsection{The 1-D frustrated spin chain}
\begin{figure}[b]
\unitlength=1mm
\centerline{\includegraphics[scale=1.0]{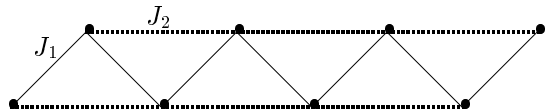}}
\caption{The non-dimerized frustrated 1-D chain.}
\label{fig2}
\end{figure}

The 1-D spin anti-ferromagnetic (AF) chain with $J_1$ spin couplings between nearest neighbor sites
and $J_2$ couplings (also AF) between next-nearest neighbor sites (cf. Fig. \ref{fig2}), is ruled by the Heisenberg 
Hamiltonian
\begin{equation}
H= 2J_1\sum_i \overrightarrow{S}_i\overrightarrow{S}_{i+1}+2J_2\sum_i \overrightarrow{S}_i\overrightarrow{S}_{i+2}.
\end{equation}
It presents a second order phase transition for $(J_2/J_1)_c=j_c=0.2411$.\cite{Ref17,Ref18,Ref19} 
\begin{figure}[t]
\centerline{\includegraphics[scale=0.38]{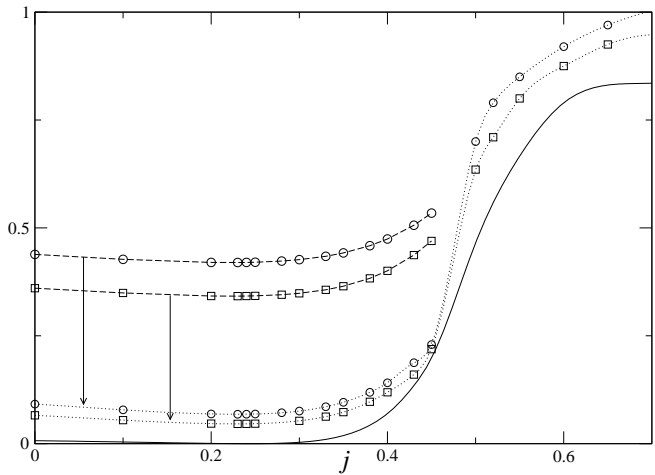}}
\caption{Dependence of the gap of the frustrated 1-D chain on the $j=J_2/J_1$ ratio. 
$--\circ--$ and $--\Box--$ direct gaps from 16 and 20 sites segments, 
$\cdots \circ \cdots$ and $\cdots \Box \cdots$ REM gaps from 8 and 10 sites 
blocks. The arrows indicate the benefit of the REM treatment. The full line gives the extrapolated gap from 
REM.}
\label{fig3}
\end{figure}
There is no gap for $J_2/J_1=j<j_c$ while a finite gap exists beyond this critical ratio. Close to the critical point 
the gap increases very slowly, presenting as \textit{essential singularity} at $j_c$.  
It behaves at this origin\cite{Ref19} as 
\begin{equation}
\Delta \simeq \beta \exp \frac{-\alpha}{j-j_c}.
\label{eq50}
\end{equation}
DMRG calculations have been reported for this system,\cite{Ref19} as well as analytic treatments.\cite{Ref20} 
The renormalized excitonic method has been applied to ($n=4, \ 6, \ 8, \ \mbox{and} \ 10$ sites) blocks, and extrapolated. 
For a given value of $n$, the calculated
gap $\Delta^{\ast\infty}$ for the lattice, estimated from Eq. \ref{eq27}, using the $\Delta^{\ast} (n)$ and 
$\Delta^{\ast} (2n)$
excitation energies, is dramatically reduced with respect to $\Delta^{\ast} (2n)$, due to the cancellation
of the $(n^{-1})$ components of the $\Delta^{\ast\infty}$ excitation.
Actually in such a simple problem \\
 $\Delta^{\ast\infty}=2\Delta^{\ast}(2n)-\Delta^{\ast}(n)$. 
If $\Delta^{\ast} (n)=A+Bn^{-1}+Cn^{-2}$
\begin{equation}
\Delta^{\ast\infty}(n)= A+C(\frac{1}{4n^2}-\frac{1}{16n^2})+\cdots =A+\frac{3C}{16n^2}.
\end{equation}
From the different calculations of $\Delta^{\ast\infty}(n)$ it is possible to estimate an extrapolated 
value of the gap. We have used a polynomial fit 
\begin{equation}
\Delta^{\ast\infty}(n)=a_1+a_2(n+1)^{-2}+a_3(n+1)^{-3}+a_4(n+1)^{-4},
\end{equation}
which gives the results reported in Fig. \ref{fig3}. One may notice that
\begin{itemize}
\item[-] the extrapolated value of the gap $\Delta^{\ast\infty}=a_1$ for $j<j_c$ is not strictly zero. The 
largest error is for $j = 0$ where $\Delta^{\ast\infty}=0.0068J_1$. This value is within the accuracy 
of the extrapolation techniques of DMRG (cf. Fig. \ref{fig3} of Ref. 19 ),
\item[-] the calculated gap goes through a minimum at $j=0.24$, close to the critical value, where 
$\vert \Delta^{\ast\infty}\vert=3.10^{-6}$
\item[-] it increases for larger values of $J_2$. Immediately beyond $J_{2c}$ the gap follows the 
expected law. We found 
\begin{eqnarray*}
\alpha & = & 0.21022 \\
\beta & = & 0.10253
\end{eqnarray*}
for the parameters of Eq. \ref{eq50} 
\item[-] the calculated gap for the Majumdar-Ghosh point ($2J_2=J_1$) is $0.465J_1$, which compares well 
with the DMRG\cite{Ref19} estimate ($0.48J_1$) and the result of an analytic development 
($\simeq 0.45J_1$)\cite{Ref20}
\end{itemize}
\subsection{The 1/5-depleted square 2-D lattice}
\begin{figure}[t]
\includegraphics[scale=1.0]{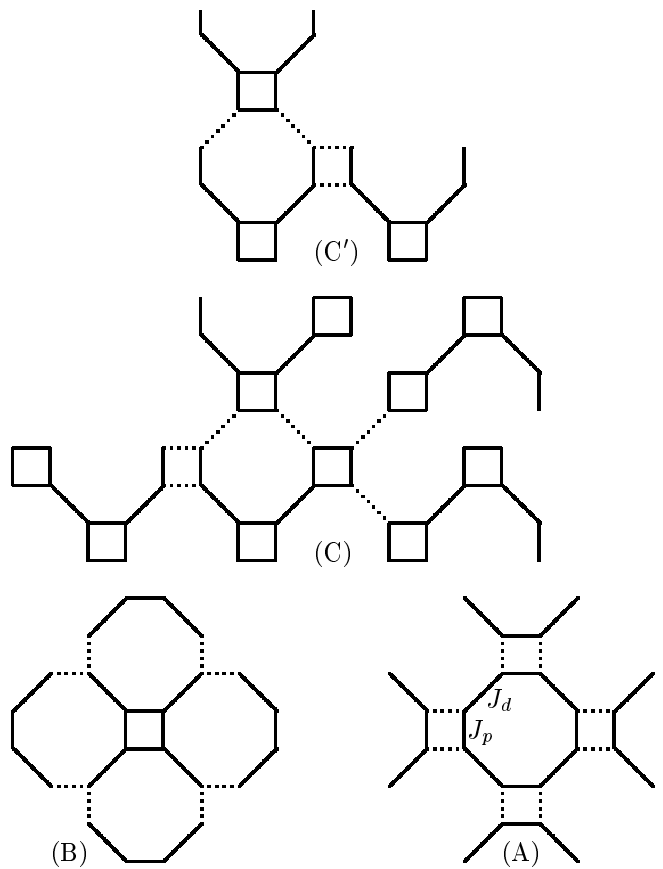}
\caption{1/5-depleted 2-D square lattice, definition of various blocks.}
\label{fig4}
\end{figure}

\begin{figure}[t]
\includegraphics[scale=0.38]{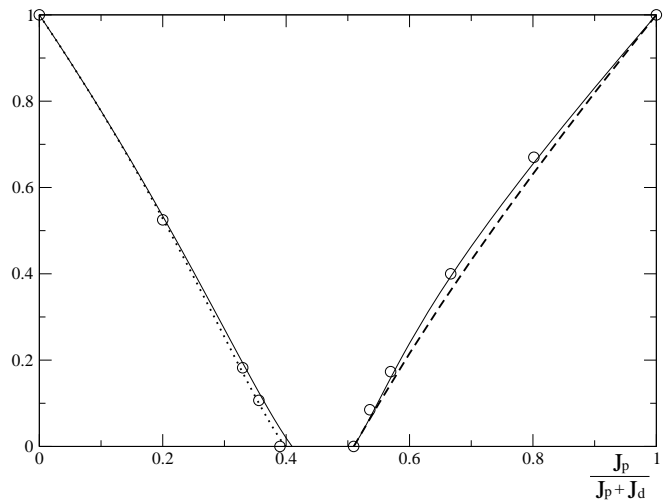}
\caption{Gap in the 1/5-depleted 2-D square lattice. (---) from octagonal blocks (A), (...) from 
blocks of type (B), (- -) from blocks of type ($\mbox{C}'$) and (C), after extrapolation,( $\circ$) QMC 
calculations from Ref. 22.}
\label{fig5}
\end{figure}

The 1/5-depleted square 2-D lattice, built from square plaquettes and octagons (cf. Fig. \ref{fig4}), was 
first considered as representing the 2-D lattice of the $CaV_4O_9$ crystal. It appeared later on that 
next-nearest neighbor spin couplings are important in this material, but the simple picture, with $J_p$ AF 
couplings on plaquette bonds and $J_d$ AF couplings between adjacent plaquettes already presents an 
interesting physics with three phases. When the plaquettes are weakly coupled, i.e., $J_p/(J_p+J_d)=j>j_c$, the 
system, in this plaquette phase, is gapped. It is also gapped when the dimers connecting the plaquettes 
are weakly coupled, i.e., when $j<j'_c$. 
This phase is called dimer-phase. In between,
i.e., for $j_c<J_p/(J_p+J_d)<j'_c$, the lattice keeps a N\'eel order and this phase is gapless. Several 
studies, using perturbative expansions\cite{Ref21} or quantum Monte Carlo (QMC) calculations \cite{Ref22} agree on this 
picture and propose $j_c \simeq 0.4\pm0.01$ and $j'_c \simeq 0.51\pm0.01$. We have tested our method on this 
problem. The simplest block that one may consider is the octagon (see schema (A) of Fig. \ref{fig4}). 
It is non-degenerate whatever the $J_p/J_d$ ratio. Actually starting from these blocks, REM provides a 
correct picture of the physics, since the gap disappears between $j_c=0.40967$ and $j'_c=0.50945$ (cf. Fig. \ref{fig5}).
This result is obtained from 8 sites blocks. 

In order to ckeck wether this excellent agreement was not fortuitous we have introduced next-nearest neighbor 
interactions between octagons, applying the formalism of section II C. Two types of trimers (linear and perpendicular) 
have to be considered. The results appear in Table 1, and they deserve the following comments 
\begin{table}[t]
\caption{Calculated spin gap for the 1/5-depleted square lattice as a function of $j=J_p/(J_p+J_d)$}
\begin{ruledtabular}
\begin{tabular}{ccccc}
      &          &          & \multicolumn{2}{c}{number of the third} \\ 
      & \multicolumn{2}{c}{\raisebox{1.5ex}[0cm][0cm]{gap}}  & \multicolumn{2}{c}{relevant eigenvector} \\
\hline
$j$   & dimers   & trimers  & linear& perpendicular \\
\hline
0     & 1.0      & 1.0      & 3     & 3  \\
0.1   & 0.777541 & 0.794138 & 3     & 3  \\
0.2   & 0.533681 & 0.561073 & 3     & 3  \\
0.3   & 0.272110 & 0.289024 & 3     & 3  \\
0.39  & 0.041669 & 0.015530 & 3     & 3  \\
0.39572 &0.028882& 0.0      & 3     & 3  \\
0.4   & 0.019637 & -0.01160 & 3     & 3  \\
0.40967 & 0.0    & -0.03536 & 4     & 3  \\
0.41  & -0.00068 & -0.03610 & 4     & 3  \\
0.42  & -0.01882 & -0.05463 & 4     & 4  \\
0.49  & -0.03646 & -0.02158 & 4     & 4  \\
0.49784 &-0.02361& 0.0      & 4     & 4  \\   
0.5   & -0.01961 & 0.005945 & 4     & 4  \\
0.50945 & 0.0    & 0.027081 & 4     & 3  \\
0.51  & 0.001129 & 0.028707 & 4     & 3  \\
0.52  & 0.024859 & 0.057802 & 4     & 3  \\
0.55  & 0.105265 & 0.139275 & 3     & 3  \\
0.7   & 0.463095 & 0.461658 & 3     & 3  \\
0.8   & 0.654426 & 0.648198 & 3     & 3  \\
0.9   & 0.831083 & 0.830046 & 3     & 3  \\
1     & 1.0      & 1.0      & 3     & 3  \\
\end{tabular}
\end{ruledtabular}
\end{table}
\begin{itemize}
\item[-] the dependence of the gap on the $j$ ratio is almost the same as when working with dimers only. The gapless
domain in slightly reduced to the interval $0.39572 <j<0.49784$
\item[-] the 3rd target vector for the perpendicular trimer (i.e., th 3rd vector presenting
the largest projection on the model space) is the 3rd eigenvector ($\Psi_3$) of the perpendicular trimer 
problem for $j\leq 0.41$ and the 4th ($\Psi_4$) one for $j>0.41$.
This may be seen as a signature for a finite (24 sites) cluster of the vicinity of the phase transition in the periodic
lattice. A similar phenomenon is observed for the linear trimer between $j=0.40$ and $j=0.41$. Regarding the
N\'eel-plaquette phase transition, a similar change of the target vectors appears for $0.50 < j < 0.51$ in the
perpendicular trimer superblock.
This phenomenon of discontinuity of $H^{eff}$ does not appear 
when working with dimers only. One might eventually circumvent this problem by taking a weighted energy for the third 
"root" appearing in the spectral definition of $H^{eff}$. If $P_0$ is te projector on the model space 
$\alpha = \vert \vert P_0 \Psi_3 \vert \vert = \langle P_0 \Psi_3 \vert P_0 \Psi_3 \rangle$, 
$\beta = \vert \vert P_0 \Psi_4 \vert \vert = \langle P_0 \Psi_4 \vert P_0 \Psi_4 \rangle$
and if
$H_{ABC}\vert \Psi_3 \rangle = E_3 \Psi_3 \rangle$, $H_{ABC}\vert \Psi_4 \rangle = E_4 \Psi_4 \rangle$,
one might define
$E^{\ast''}_{ABC}=\alpha E_3+ \beta E_4$, i.e. proceed to a diabatization of the 3rd target vector.
\end{itemize}
One may alternatively change the shape of the blocks. A stared 8 sites
block with 4 dimer bonds around a plaquette pictured in scheme (B) of Fig. \ref{fig4} is expected to be relevant 
for the dimer phase. 
The gap calculated from these blocks almost coincides with the previously calculated one, with a critical value 
$j_c=0.40$ (see Fig. \ref{fig5}).

For the plaquette phase we have considered blocks with one or two plaquettes, all atoms belonging to
plaquette bonds. The number of sites are 8 and 10 respectively, as pictured in schemes ($\mbox{C}'$) and (C) of
Fig. \ref{fig4}. An extrapolation leads to the curve
on the right part of Fig. \ref{fig5}. The gap is slightly smaller than from the octagons, but the critical values
of disappearance of the gap, $j'_c=0.51$, coincide. We have compared our calculated gaps with the ones
reported (in Figure 2 of Ref. 22 ) from QMC calculations, and the two methods
practically coincide (within the uncertainties of reading of the above-mentioned Figure).
\subsection{The plaquette lattice}
\begin{figure}[t]
\includegraphics[scale=1.0]{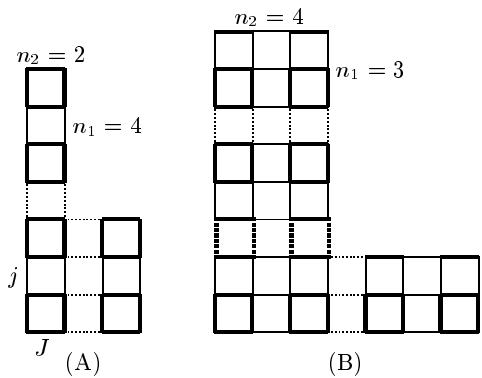}
\caption{Definitions of blocks for the study of the plaquette lattice.}
\label{fig6}
\end{figure}

The square type lattice built from interacting phaquettes is characterized by intra-plaquette $J$ and  
inter-plaquette $j$ AF couplings (cf. Fig. \ref{fig6}). The properties depend on the $\lambda =j/J$ ratio.
For $j=0$, the plaquettes are independent and the lattice is gapped. It is not gapped for the $j=J$ 
2-D square lattice, and phase transitions as expected to occur for $(j/J)_c=\lambda_c$ and for $(J/j)_c=1/\lambda_c$
(this last relation being due to the intrinsic symmetry between $j$ and $J$).
Several works have been devoted to this problem. Third order series expansions\cite{Ref23} and QMC 
calculations\cite{Ref24} suggest that $\lambda_c \simeq 0.55$. Extrapolations of finite size exact 
diagonalizations\cite{Ref25} fail to give a zero spin-gap whatever the value of $\lambda$. A recent work has 
used the CORE method\cite{Ref26} together with order parameter susceptibilities, suggesting a critical 
behavior between $\lambda = 0.5$ and $\lambda = 0.6$. The problem of the gap is reexamined here using REM.
 
\begin{figure}[t]
\centerline{\includegraphics[scale=0.38]{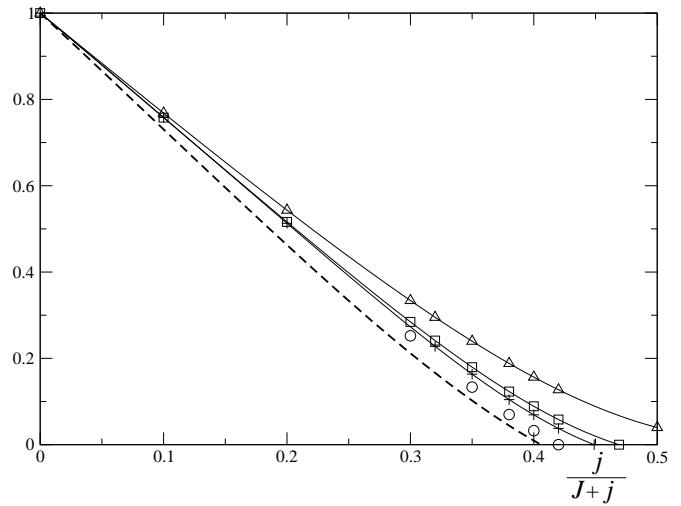}}
\caption{Singlet-triplet spin gap in the plaquette lattice from $2 \times 2$ block ($\triangle$),
$4 \times 2$ block  ($\Box$), $6 \times 2$ block (+), $4\times 3$ block ($\circ$). Dashed line: extrapolation from
$n\times 2$ block.}
\label{fig7}
\end{figure}

Two types of blocks have been considerd.\\ 
The first one involves one, two or three plaquettes ($n=4, \ 8 \ \mbox{and} \ 12$ sites),
fragments of a ladder (see schema (A) of Fig. \ref{fig6}). There are two types of dimers, collinear or side by side. 
If one calls $\Delta E_{AB}$ and $\Delta E'_{AB}$ the excitation energies for these dimers, $\Delta E_{A}$ the excitation
energy of the block, our model leads to the following expression of the gap $\Delta E(n_1,n_2)$, for a block
of $n_1$ sites along the longitudinal direction, $n_2$ sites along the transverse one,
\begin{eqnarray}
\Delta E(n_1,n_2) &=& 2\Delta E_{AB}(2n_1,n_2)+2\Delta E_{AB}(n_1,2n_2) \nonumber \\
& & -3\Delta E_{A}(n_1,n_2).
\end{eqnarray}
Fig. \ref{fig7} reports the gap calculated for $n_1=2,3,4$ and $n_2=2$. One sees that the gap vanishes when $\lambda$
tends to 1. An extrapolation is possible in terms of $n_1^{-1}$ and $n_1^{-2}$, for $n_2=2$,
\begin{equation} 
\Delta E(n_1,n_2=cte)=A_0+\frac{A_1}{n_1}+\frac{A_2}{n_1^{2}}
\label{eq54}
\end{equation}
$\overline{\Delta} E(n_2)=A_0$ is the $n_1$ extrapolated value of the gap for a fixed value of $n_2$. Assuming that
\begin{equation}
\Delta E(n_1,n_2)= (\alpha+\frac{\beta}{n_1}+\frac{\gamma}{n_1^{2}}+\cdots)
(\alpha+\frac{\beta}{n_2}+\frac{\gamma}{n_2^{2}}\cdots)
\end{equation}
\begin{equation}
\Delta E(n_1,n_2)= \overline{\Delta} E+\frac{a_1}{n_1}+\frac{a_1}{n_2}+\frac{a_2}{n_1^{2}}+\frac{a_2}{n_2^{2}}+
\frac{b}{n_1n_2}
\end{equation}
one obtains
\begin{equation}
\Delta E(n_1,n_2)= \overline{\Delta} E-\frac{a_2}{2n_1^{2}}-\frac{a_2}{2n_2^{2}}-\frac{b}{n_1n_2}.
\label{eq57}
\end{equation}
Confronting Eq. \ref{eq54} and Eq. \ref{eq57} gives $A_0=\overline{\Delta} E-a_2/2n_2^{2}$, $A_1=-b/n_2$ and $A_2=-a_2/2$. 
Hence the final value of the extrapolation gap is $\overline{\Delta} E=A_0-A_2/n_2^{2}$.
Fig. \ref{fig7} reports the so-calculated gap as well as the values $\Delta E(n_1,n_2)$ for $n_2=2$ 
and $n_1=2,4,6$ as a function of the $j/(J+j)$ ratio. One sees that a gapless phase appears for $n_1>2$. After extrapolation the lattice is found to be gapless for $j/J>2/3=0.666.$
The value of $\lambda_c$ is somewhat larger than the commonly accepted value but it represents a considerable
improvement over the extrapolations of finite lattices. One may mention that, as a by-product of the present
calculations, one obtains, for $j=J$, a value of the gap of the two-leg ladder. The extrapolation leads to
$\Delta E=0.47J$, close to the best QMC estimate ($0.50J$).\cite{Ref27} \\
A second type of rectangular blocks have been
considered, involving odd numbers of sites in one direction and even numbers in the other one 
(see schema (B) of Fig. \ref{fig6}).
In such a case there are three types of dimers. These 12 sites ($n_1=3,n_2=4$) blocks are more compact than the
previous ($n_1=6,n_2=2$) ones and the calculated gap, which appears in Fig. \ref{fig7}, is somewhat lower. Extrapolation
is difficult in this case, due to the difference in the physical nature of the dimers, but the evaluations
from different blocks are quite consistent.
\section{Conclusion}
We have presented a very simple method for the study of the gap in gapped periodic lattices. The method
rests on the consideration of blocks and a truncation of the Hilbert
space to products of a few eigenstates of the block as practiced in the RSRG. 
In the past we have considered $(2n+1)$ sites blocks, with spatially non-degenerate doublet ground states,
in spin lattices. The blocks can then be seen as $S_z=\pm1/2$ quasi-spins. Using the theory of effective
Hamiltonians, and the exact spectrum of dimers (or trimers) of blocks, we have proposed to renormalize
the interactions between blocks, and the so-obtained variant of CORE (RSRG-EI)
happens to keep the conceptual elegance of Wilson's idea while gaining, at a very low cost, numerical 
relevance.\cite{Ref7,Ref8} 

The present work is closely related but different. It considers blocks with even number of sites, presenting a 
non-degenerate ground state. Again the exact treatment of the block and of the dimers or trimers of blocks is 
employed to define block effective energies and inter-block effective interactions. 
However different model spaces are used for the ground state and for the lowest excited states.
For the ground state $\Psi_0$, built from the product of ground states, the energy is a simple sum of intra and
inter-block energies. 
The excited states are linear combinations of locally singly excited functions, products of an excited state on  
one block by the ground states functions on the other blocks. This space is a small fraction of those handled in RSRG 
techniques.
The knowledge of the excited states of dimers or trimers of blocks enables one to define the effective interactions 
between an excited block and its ground state neighbors, as well as effective excitation hopping integrals, which 
delocalize the excitations. The effective interactions incorporate complex processes, including multiple 
excitations or/and inter-block excitations. These informations are used to build an excitonic Hamiltonian for the 
infinite lattice, and to estimate the gap. 

The method does not provide any information on the low energy physics of gapless phases but solves some dramatic failures 
of the use of bare interactions (as they manifest in the original RSRG formalism). It can be applied to various 
Hamiltonians (Tight-binding, Hubbard, Heisenberg, ab-initio). 
The method has been presented (and tested) in its simplest version on spin lattices, with identical blocks, 
one excited state per block, and extraction from dimers and trimers. It is possible to generalize it to blocks of 
different sizes or topologies, and one may keep several excited states per block.
The renormalized excitonic method has been tested so far to the research of singlet-triplet gaps but it is
applicable as well to singlet to singlet excitations. The bottleneck is the size of 
the dimers or trimers of blocks, the lowest states of which have to be calculated. In the few benchmark 
problems tested in the present work the results are surprisingly accurate and the method seems to be
able to locate phase transitions between gapped and gapless phases in 1-D and 2-D lattices at a very low computational cost.
An other application concerning the Shastry-Sutherland,\cite{Ref28} shows the relevance the here proposed method for the 
study of phase transition in frustrated 2-D spin lattices.
\begin{acknowledgments}
The authors thank R. Bastardis for his help and S. Capponi and M. Mambrini for stimulating discussions.
\end{acknowledgments}

\end{document}